\documentclass[12pt]{article}
\usepackage{verbatim,color,amssymb,epsfig}
\usepackage[usenames,dvipsnames]{xcolor}
\usepackage{graphicx}
\usepackage{subfigure}
\usepackage{alg}
\usepackage{fancyhdr}

\usepackage{amsmath}
\usepackage{natbib}
\usepackage{enumitem}

\def\pr{\hbox{pr}}

\renewcommand{\baselinestretch}{1.5}

\begin{document}

\def\x{{\mathbf x}}
\def\L{{\cal L}}

\newcommand{\wt}[1]{\tilde{#1}}
\newcommand{\mc}[1]{\mathcal{#1}}
\newcommand{\bigO}{\mathcal{O}}
\newcommand{\A}{\mathcal{A}}
\newcommand{\R}{\mathbb{R}}
\newcommand{\C}{\mathbb{C}}
\newcommand{\Z}{\mathbb{Z}}
\newcommand{\N}{\mathbb{N}}
\newcommand{\E}{\mathbb{E}}
\renewcommand{\P}{\mathbb{P}}
\newcommand{\sA}{\mathscr{A}}
\newcommand{\sM}{\mathscr{M}}
\newcommand{\sY}{\mathscr{Y}}
\newcommand{\CD}{\mathcal{D}}
\newcommand{\CT}{\mathcal{T}}
\newcommand{\CA}{\mathcal{A}}
\newcommand{\CV}{\mathcal{V}}
\newcommand{\CH}{\mathcal{H}}

\newcommand{\notate}[1]{\textcolor{red}{\textbf{[#1]}}}

\makeatletter
\def\algspacing{\alg@unmargin}
\makeatother

\newcommand{\algfor}[1]{\textbf{for} #1\\\algbegin}
\newlength{\algorithmwidth}
\algorithmwidth=0.98\textwidth

\newcommand{\sG}{\mathscr{G}}
\newcommand{\sd}{\Sigma\Delta}
\newcommand{\psum}{\mathop{\sum\nolimits'}}
\newcommand{\pprod}{\mathop{\prod\nolimits'}}
\newcommand{\diag}{\operatorname{diag}}
 \newcommand{\scalprod}[2]{\left\langle #1,#2 \right\rangle}

\newtheorem{theorem}{\underline{\bf Theorem}}
\newtheorem{Assume}{\underline{\bf Assumptions}}
\newtheorem{proposition}{\underline{\bf Proposition}}
\newtheorem{definition}{\underline{\bf Definition}}
\newtheorem{remark}{\underline{\bf Remark}}
\newtheorem{lemma}{\underline{\bf Lemma}}
\newtheorem{cor}{\underline{\bf Corollary}}
\newtheorem{problem}{\underline{\bf Problem}}
\newtheorem{Eq}{Equation}

\setlength{\textheight}{9in}
\setlength{\textwidth}{6.5in}
\setlength{\topmargin}{-36pt}
\setlength{\oddsidemargin}{0pt}
\setlength{\evensidemargin}{0pt}
\tolerance=500
\renewcommand{\baselinestretch}{1.5}
\def\boxit#1{\vbox{\hrule\hbox{\vrule\kern6pt
          \vbox{\kern6pt#1\kern6pt}\kern6pt\vrule}\hrule}}
\def\rjccomment#1{\vskip 2mm\boxit{\vskip 2mm{\color{black}\bf#1} {\color{blue}\bf -- RJC\vskip 2mm}}\vskip 2mm}

\thispagestyle{empty}
\baselineskip=28pt
{\LARGE{\bf  Testing Hardy-Weinberg equilibrium with a simple root-mean-square statistic }}

\begin{center}
Rachel Ward\footnote{\baselineskip=10pt Department of Mathematics, RLM 10.144, University of Texas at Austin, 2515 Speedway, Austin, TX 78712, U.S.A., rward@math.utexas.edu.  Supported in part by a Donald D. Harrington Faculty Fellowship, Alfred P. Sloan Research Fellowship, and DOD-Navy grant N00014-12-1-0743.} and
Raymond J. Carroll\footnote{\baselineskip=10ptDepartment of Statistics, 3143 TAMU, Texas A$\&$M University, College Station, TX 77843, U.S.A., carroll@stat.tamu.edu. Research supported by a grant from the National Cancer Institute (R37-CA057030). This publication is based in part on work supported by Award Number KUS-CI-016-04, made by King Abdullah University of Science and Technology (KAUST).}
\end{center}

\begin{center}
{\Large{\bf Abstract}}
\end{center}
\baselineskip=12pt
We provide evidence that, in certain circumstances, a root-mean-square test of goodness-of-fit can be significantly more powerful than state-of-the-art tests in detecting deviations from Hardy-Weinberg equilibrium.   Unlike Pearson's chi-square test, the log--likelihood-ratio test, and Fisher's exact test, which are sensitive to \emph{relative} discrepancies between genotypic frequencies, the root-mean-square test is sensitive to \emph{absolute} discrepancies.  This can increase statistical power, as we demonstrate using benchmark data sets and simulations, and through asymptotic analysis.  

\baselineskip=12pt
\par\vfill\noindent
\underline{\bf Some Key Words}:
Chi-square test; Common genotypes; Exact tests; Goodness-of-fit; Hardy-Weinberg equilibrium; p-value; Power; Root-mean-square.

\par\medskip\noindent
\underline{\bf Short title}: Hardy-Weinberg equilibrium

\clearpage\pagebreak\newpage
\pagenumbering{arabic}
\newlength{\gnat}
\setlength{\gnat}{22pt}
\baselineskip=\gnat

\section{Introduction}
In 1908, G. H. Hardy \citep{hardy} and W. Weinberg \citep{weinberg} independently derived mathematical equations to corroborate the theory of Mendelian inheritance, proving that in a large population of individuals subject to random mating, the proportions of alleles and genotypes at a locus stay unchanged unless specific disturbing influences are introduced.  Today, Hardy-Weinberg equilibrium (HWE) is a common hypothesis used in scientific domains ranging from botany \citep{weising05} to forensic science \citep{dna} and genetic epidemiology \citep{sham01, klb04}.  Statistical tests of deviation from Hardy-Weinberg equilibrium are fundamental for validating  such assumptions.   Traditionally, Pearson's chi-square goodness-of-fit test, or an asymptotically-equivalent variant such as the log--likelihood-ratio test, was used for this assessment.   Before computers became readily available, the asymptotic chi-square approximation for the statistics used in these tests, however poor, was the only practical means for drawing inference.   With the now widespread availability of computers, exact tests can be computed effortlessly, opening the door to more powerful goodness-of-fit tests.  In their seminal paper, \cite{gt}  campaigned for an exact test of HWE based on the likelihood function.  While their work renewed interest in conditional exact tests for Hardy-Weinberg equilibrium \citep{rr94, diaconis98, wca05}, likelihood-based tests have also been subject to criticism, and there is little evidence that such tests are more powerful than other exact tests, such as those based on likelihood-ratios \citep{engels09} or the root-mean-square.

In this article, we demonstrate, using the classical data sets from \cite{gt} and several numerical experiments, that goodness-of-fit tests based on the root-mean-square distance can be up to an order-of-magnitude more powerful than all of the classic tests at detecting meaningful deviations from Hardy-Weinberg equilibrium.  The classic tests, tuned to detect \emph{relative} discrepancies, can be blind to overwhelmingly large discrepancies among common genotypes that are drowned out by expected finite-sample size fluctuations in rare genotypes. The root-mean-square statistic, on the other hand, is tuned to detect deviations in \emph{absolute} discrepancies, and easily detects large discrepancies in common genotypes.

None of the statistics we consider produces a test that is uniformly more powerful than any other.  At the very least, the root-mean-square statistic and the classic statistics focus on complementary classes of alternatives, and their combined p-values provide a more informative test than either p-value used on its own.

The results of our analysis are consistent with the numerous experiments conducted in recent work  \citep{ptw3}, which highlight the power of the root-mean-square statistic over classic statistics in detecting meaningful discrepancies in nonuniform distributions.  \cite{tygert12} provides several representative examples for which the root-mean-square test is more powerful than Fisher's exact test for homogeneity in contingency-tables.

This article is structured as follows: in Section 2 we recall the set-up and motivation for testing Hardy-Weinberg equilibrium.  We describe the relevant test statistics in Section 3, and in Section 4 we compare the performance of these statistics on the classic data sets from Guo and Thompson, and also compare the  power and Type I error of the statistics in detecting deviations due to inbreeding and selection.
We provide an asymptotic analysis of the various statistics in Section 5 to highlight the limited power of the classic statistics compared to the root-mean-square statistic in distinguishing important classes of deviations from Hardy-Weinberg equilibrium, and end with concluding remarks in Section 6. {\bf Supplementary Material} includes pseudocode for algorithms and proofs of technical results.

\section{Hardy-Weinberg equilibrium: set-up and motivation}\label{sec:setup}

\begin{figure}[h]
\centering
\includegraphics[width=6cm]{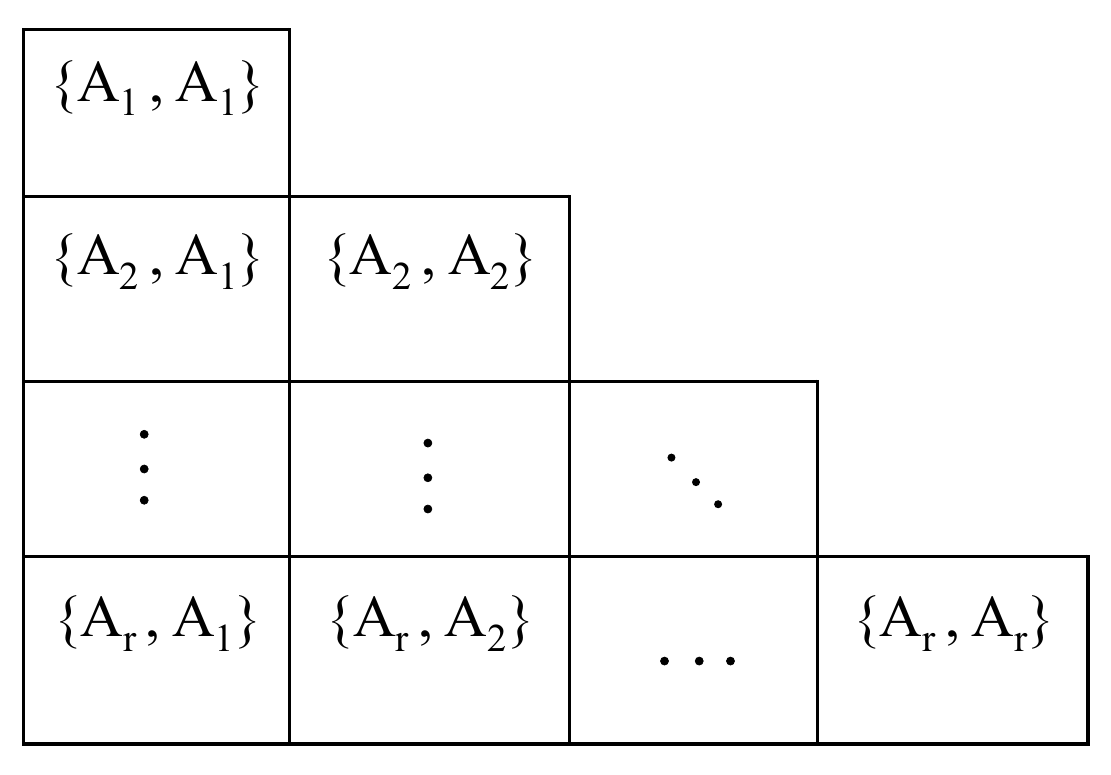}
\caption{\baselineskip=10pt Enumeration of genotypes for a gene having $r$ alleles $\textrm{A}_1,\textrm{A}_2, \dots,\textrm{A}_r$.}
\label{fig1}
\end{figure}

Recall that a \emph{gene} refers to a segment of DNA at a particular location (locus) on a chromosome.  The gene may assume one of several discrete variations, and these variants are referred to as \emph{alleles}.  An individual carries two alleles for each autosomal gene --- one allele selected at random from the pair of alleles carried by the mother, and one allele selected at random from the pair of alleles carried by the father.  These two alleles, considered as an unordered pair, constitute the individual's \emph{genotype}.  A gene having $r$ alleles $\textrm{A}_1,\textrm{A}_2, \dots,\textrm{A}_r$ has $r(r+1)/2$ possible genotypes.  These genotypes are naturally indexed over a lower-triangular array as in Figure \ref{fig1}.

\begin{figure}
\centering
\subfigure[{\bf Example 1:} $n = 45$.]{\includegraphics[width=4.5cm]{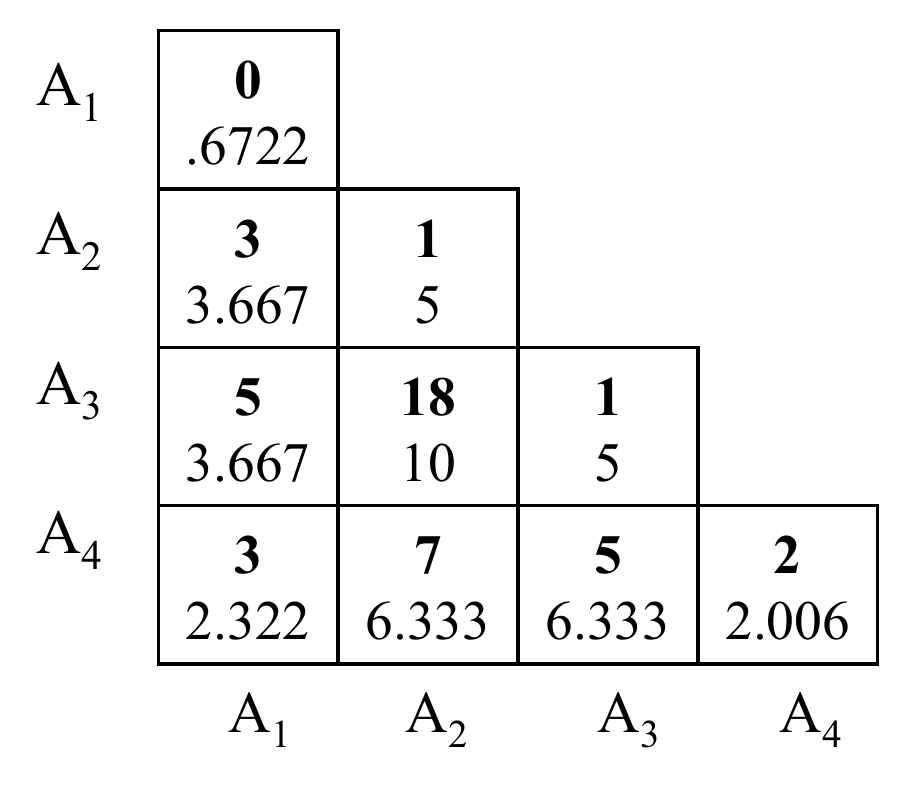}}
\quad \quad  \quad
\subfigure[{\bf Example 2:} $n = 8297$.]{\includegraphics[height=8cm,width=10cm]{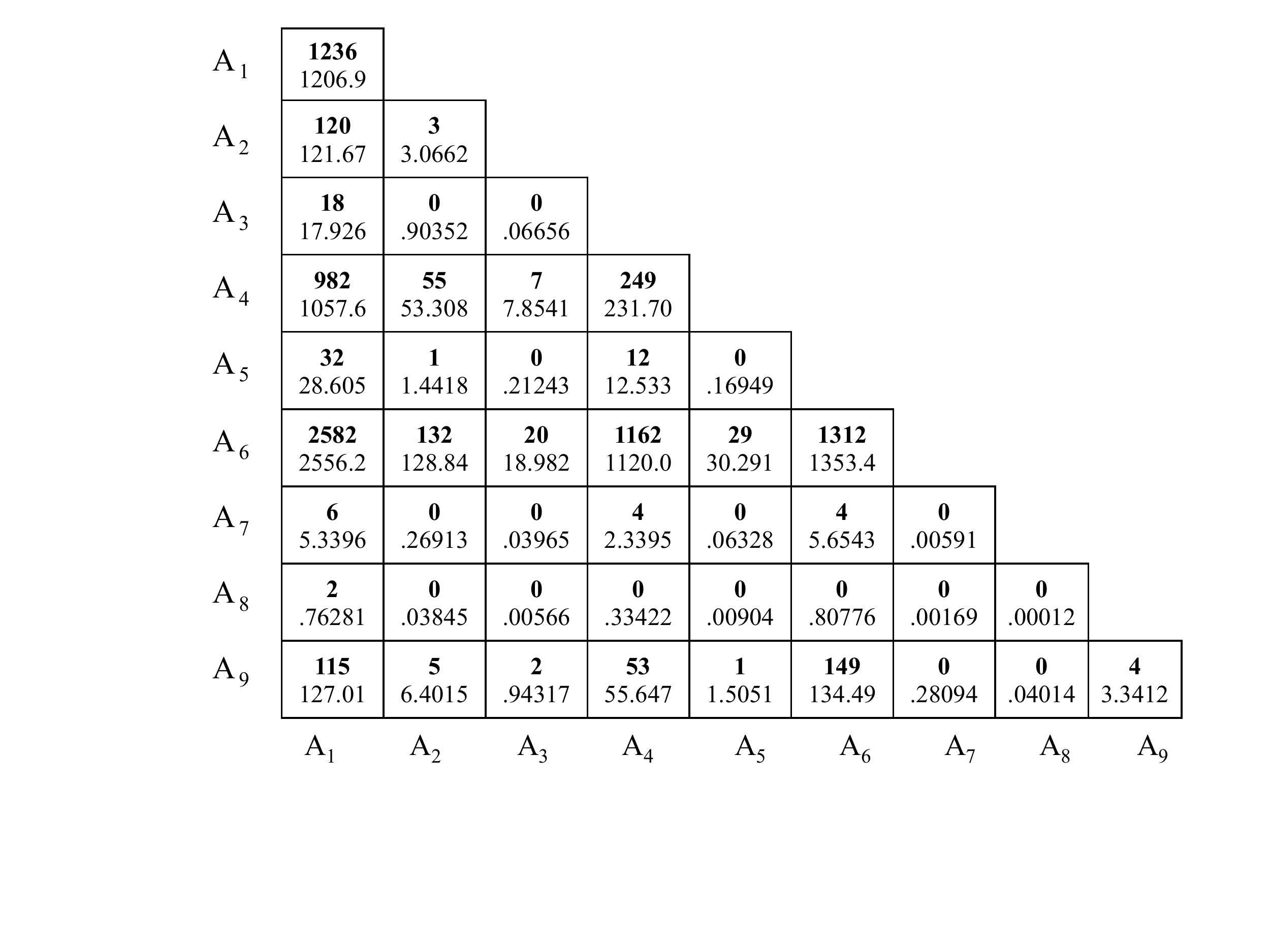}}
\vspace{5mm}
\subfigure[{\bf Example 3:} $n = 30$.]{\includegraphics[width=9cm,height=8cm]{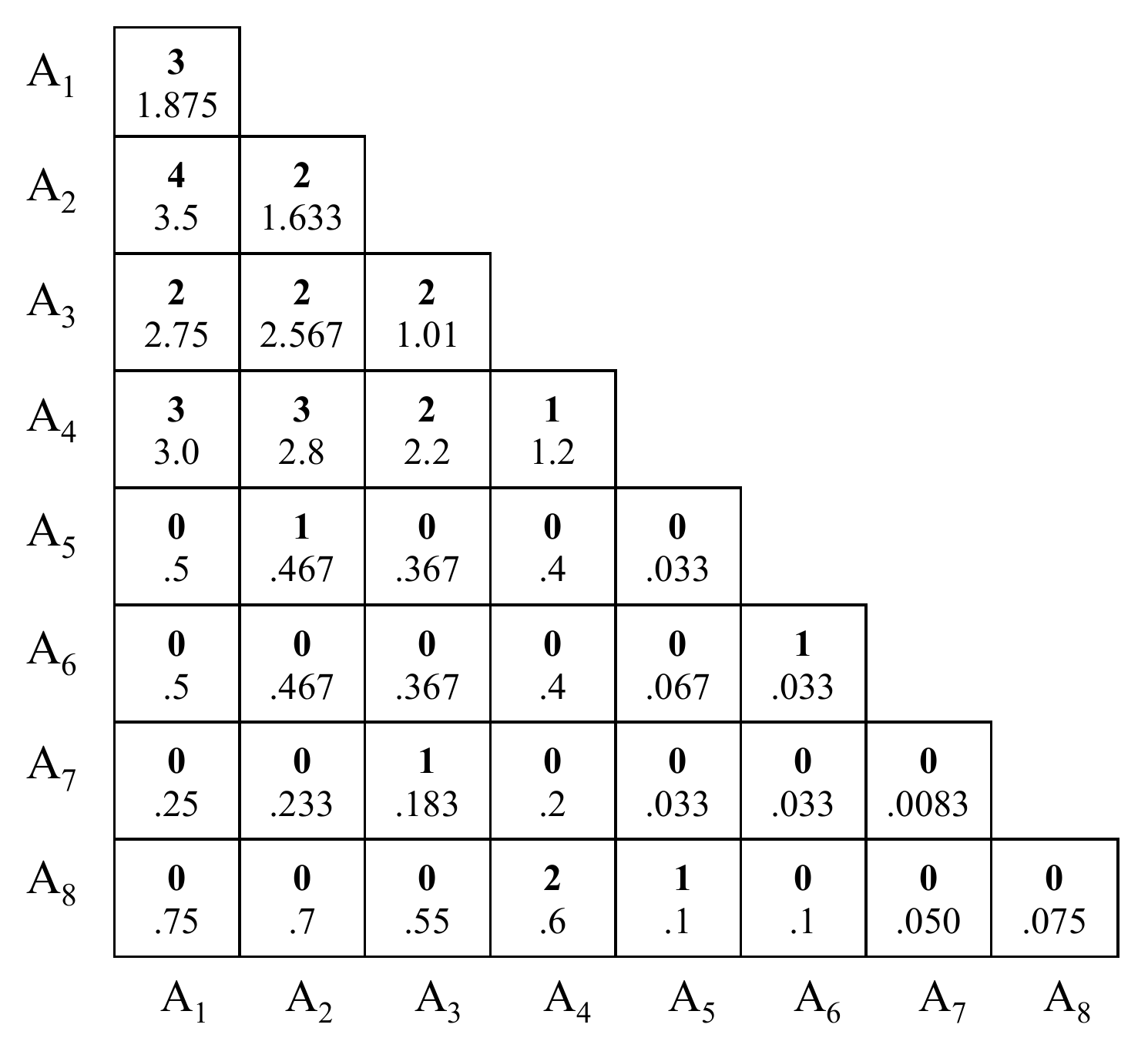}}
\caption[]{\baselineskip=12pt The three data sets from \cite{gt}. Observed counts are in bold and expected counts under HWE are below.}
\label{examples}
\end{figure}

A population is said to be in \emph{Hardy-Weinberg Equilibrium} (HWE) if the following holds. If $p_{j,k}$ is the relative proportion of genotype $\{\textrm{A}_j,\textrm{A}_k\}$ in the population, and if $\theta_k$ is the proportion of allele $A_k$ in the population, then the system is in HWE if
\begin{equation}
\label{hwe1}
p_{j,k} = p_{j,k}(\theta_j, \theta_k) =
\left\{ \begin{array}{cc}2\theta_j \theta_k, & j > k \\
          \theta_k^2, & j = k.
  \end{array} \right.
\end{equation}

\section{Testing Hardy-Weinberg equilibrium}
A random sample of $n$ genotypes $X_1, X_2, \dots X_n$ from this population can be regarded as a sequence of independent and identical draws from the multinomial distribution specified by probabilities
\begin{equation}
\label{multi}
\pr( X_{i} = \{\textrm{A}_j,\textrm{A}_k \}) = p_{j,k}, \quad 1 \leq k \leq j \leq r.
\end{equation}
If $n_{j,k}$ realizations of genotype $\{\textrm{A}_j,\textrm{A}_k\}$ are observed in the sample of $n$ genotypes, then the number of instances of allele $\textrm{A}_j$ in the observed sample of $2n$ alleles is
\begin{equation}
\label{allelecount}
n_j = \sum_{k=j}^r n_{k,j} + \sum_{k=1}^{j} n_{j,k}, \quad \quad j = 1, \dots, r.
\end{equation}
In order to gauge the consistency of the sample counts $(n_{j,k})$ with Hardy-Weinberg equilibrium, we must first specify the $r-1$ free parameters $\theta_1, \theta_2, \dots, \theta_{r-1}$ corresponding to the underlying allele proportions in the HWE model \eqref{hwe1}.  The \emph{observed} proportions of alleles, $n_1/(2n), n_2/(2n), \dots, n_{r-1}/(2n)$,  are the maximum likelihood estimates of $\theta_1, \theta_2, \dots, \theta_{r-1}$ in the family of HWE equilibrium equations \eqref{hwe1};  these parameter specifications give rise to the \emph{model counts} of genotypes under Hardy-Weinberg equilibrium,
\begin{eqnarray}
m_{jk} = (2-\delta_{jk})(n_j \  n_k)/(4n), \label{eq02}
\end{eqnarray}
where $\delta_{jk}$ is the Kronecker delta function with $\delta_{jk} = 1$ if $j=k$ and $= 0 $ otherwise.
A \emph{goodness-of-fit} test serves as an omnibus litmus test to gauge consistency of the data with HWE.  Ideally, the goodness-of-fit test should be sensitive to a wide range of possible local alternatives; more realistically, several different goodness-of-fit tests can be used jointly, each sensitive to its own class of alternatives.  If a nonparametric test as such indicates deviation from equilibrium, different parametric tests can then be used to elucidate particular effects of the deviation such as directions of disequilibrium or level of inbreeding.  Several parametric Bayesian methods have been proposed as well \citep{chenthomson, hw97, ab97, hw08, hw9, hw11}.  In this paper we will focus only on nonparametric (or nearly nonparametric) tests of fit, but we emphasize that goodness-of-fit tests should be combined with Bayesian approaches and other types of evidence for and against the HWE hypothesis before drawing final inference.

\subsection{Goodness-of-fit testing}

A goodness-of-fit test compares the model and empirical distributions using one of many possible measures.  Three classic measures of discrepancy, all special cases of Cressie-Read power divergences, are Pearson's $\chi^2$-divergence
 \begin{equation}
 \label{chi2}
X^2 = \sum_{1 \leq k \leq j \leq r} ( n_{j,k} - m_{j,k} )^2/m_{j,k},
 \end{equation}
the log--likelihood-ratio or $g^2$ divergence,
 \begin{equation}
 \label{g2}
G^2 =  2 \sum_{1 \leq k \leq j \leq r} n_{j,k} \log( n_{j,k}/m_{{j,k}} ),
 \end{equation}
and the Hellinger distance
\begin{equation}
\label{H2}
H^2 = 4\sum_{1 \leq k \leq j \leq r} \big(  \sqrt{n_{j,k}} - \sqrt{m_{j,k}} \big)^2.
\end{equation}
Another classic measure of discrepancy is the negative log--likelihood function, which is based directly on the likelihood function for the multinomial distribution,
\begin{eqnarray}
\label{nll}
L &=& -\log({\cal L}), \hbox{where} \\
\label{likelihood}
{\cal L}(n_{j,k}; \hspace{.5mm} n, \hspace{.5mm} m_{j,k}) &=&\textrm{Prob}\Big( N_{1,1} = n_{1,1}, \hspace{1mm} N_{2,1} = n_{2,1}, \hspace{1mm} \hdots ,N_{r,r} = n_{r,r} \Big) \nonumber \\
&=& \frac{n!}{n_{1,1}!n_{1,2}! \dots n_{r,r}!n^n} m_{1,1}^{n_{1,1}}m_{1,2}^{n_{1,2}}\dots m_{r,r}^{n_{r,r}}.
\end{eqnarray}
The negative log-likelihood statistic (\ref{nll})  looks similar to the log--likelihood-ratio statistic $G^2$, but there is an important distinction to be made: the log--likelihood-ratio, which sums the logarithms of \emph{ratios} between observed and expected counts, is a proper divergence.  The negative log--likelihood function is not a divergence, and this results in several undesirable properties that have led many to criticize its use \citep{gp75, ra75, engels09}.

The negative log--likelihood function does have something in common with the power-divergence discrepancies: under the null-hypothesis, the negative log--likelihood statistic $L$ and the power divergence statistics $X^2, G^2$, and $H^2$ all become a chi-square random variable with $r(r-1)/2-1$ degrees of freedom as the number of draws $n$ goes to infinity and number of alleles remains fixed \citep{brownlee}.  Before computers became widely available, using a statistic with known asymptotic approximation was necessary for obtaining any sort of approximate p-value.  The exact (non-asymptotic) p-values for these statistics or any other measure of discrepancy can now be computed effortlessly using Monte-Carlo simulation.

In this paper, we distinguish two types of commonly-used p-values, which we refer to as the \emph{plain} p-value and \emph{fully conditional} p-value. One could also consider Bayesian p-values \citep{gelman03}, among other formulations.

 To compute the plain p-value, one repeatedly simulates $n$ independent and identically distributed draws from the model multinomial distribution $(m_{j,k}/n)$.  For each simulation $i$, the genotype counts $N_{j,k}^{(i)}$, allelic counts $N_j^{(i)} =  \big( \sum_{k=j}^r N^{(i)}_{k,j} + \sum_{k=1}^{j} N^{(i)}_{j,k} \big),$ allelic proportions $\Theta_{j}^{(i)} = N_j^{(i)}/(2n),$ and equilibrium model counts associated to this sample, $M_{j,k}^{(i)} = (2-\delta_{j,k}) N_j^{(i)} N_k^{(i)}/(4n)$, are computed.  The plain p-value is the fraction of times the discrepancy between the simulated counts $(N^{(i)}_{j,k})$ and their model counts $(M_{j,k}^{(i)})$ is at least as large as the measured discrepancy between the observed counts $n_{j,k}$ and their model counts $m_{j,k}$. \cite{Henze} shows that this procedure has an asymptotically correct Type I error for fixed $r$ as $n \to \infty$.  This procedure for producing p-values can be viewed as a parametric bootstrap approximation, as discussed for example, by \cite{et93}, \cite{Henze}, and \cite{brs06}.

The fully conditional p-value corresponds to imposing additional restrictions on the probability space associated to the null hypothesis.   To compute the fully conditional p-value, the observed counts of alleles, $n_1, \dots, n_r$, are treated as known quantities in the model, to remain fixed upon hypothetical  repetition of the experiment.  This would hold, for example, if the sample population used in the experiment were the entire population of individuals.   More specifically, one repeatedly simulates $n$ i.i.d. draws from the \emph{hypergeometric} distribution that results from conditioning the multinomial model distribution $(m_{j,k}/n)$ on the observed allele counts, $N_1 = n_1, N_2 = n_2, \dots, N_r = n_r$.
\cite{gt} provided an efficient means for performing such a simulation: apply a random permutation to the sequence
\begin{equation}
\label{string}
{\cal A} =\Big\{  \underbrace{\overbrace{\textrm{A}_1,\textrm{A}_1, \dots,\textrm{A}_1}^{n_1}, \hspace{.5mm} \overbrace{\textrm{A}_2, \dots,\textrm{A}_2}^{n_2}, \hspace{.5mm} \dots, \overbrace{\textrm{A}_r \dots\textrm{A}_r}^{n_r} }_{2n} \Big\},
\end{equation}
and identify the pairs $\{\textrm{A}_{2j},\textrm{A}_{2j+1}\}$.  The fully conditional p-value is the fraction of times the discrepancy between the simulated counts $(N^{(i)}_{j,k})$ and the model counts $(m_{j,k})$ is at least as large as the measured discrepancy.

Pseudocode for calculating plain and fully conditional p-values is provided in Algorithms \ref{alg:1} and \ref{alg:2} of Appendix \ref{app2} in the {\bf Supplementary Material}.

\subsection{The root-mean-square statistic}
A natural measure of discrepancy for goodness-of-fit testing which has not received as much attention in the literature is the root-mean-square distance,

\begin{equation}
\label{rms}
f = \left\{\frac{2}{n^2 r(r+1)} \sum_{1 \leq k \leq j \leq r} (n_{j,k} - m_{j,k} )^2\right\}^{1/2}.
\end{equation}
In contrast to the classic statistics, the asymptotic distribution for the root-mean-square statistic $F$ in the limit of infinitely many draws and fixed alleles, while completely well-defined and efficient to compute, depends on the model distribution, as described by \cite{ptw1, ptw2}.  Using the pseudocode provided in Algorithms \ref{alg:1} and \ref{alg:2} of Appendix \ref{app2} in the {\bf Supplementary Material}, we can compute p-values for the root-mean-square statistic.

\section{Numerical results}\label{experiment}
\subsection{Benchmark data sets}\label{sec4.1}
We next compare the performances of  the root-mean-square statistic and the classic statistics in detecting deviations from Hardy-Weinberg equilibrium.  We first evaluate the performance of the various statistics on three benchmark data sets from~\cite{gt}.  The three data sets, which we refer to as Examples 1, 2, and 3, are represented in Figure \ref{examples} as lower-triangular arrays of counts.  The bold entry in each cell corresponds to the number $n_{j,k}$ of observed counts of genotype $\{\textrm{A}_j,\textrm{A}_k \}$ in the sample, and the second entry in each cell corresponds to the expected number $m_{j,k}$ of counts under HWE.

For each example, and for each of the five test statistics $X^2, G^2, H^2, L,$  and $F$, we calculate both the plain and fully conditional p-values using $16,000,000$ Monte-Carlo simulations for each calculation.  The results of the analyses of Examples 1-3 are displayed in Table 1. We next discuss the results for each example.

\subsubsection{Graphical views of the data}\label{sec4.2}

Figures \ref{fig:box2}, \ref{fig:box3}, and \ref{fig:box1} contain boxplots displaying the median, upper and lower quartiles, and whiskers reaching from the 1st to 99th percentiles for relative root-mean-square discrepancies and relative chi-square discrepancies simulated under the plain Hardy-Weinberg equilibrium null hypothesis for the data sets from Examples 1, 2, and 3. The boxplots are for simulated data, whereas the large open circles indicate the observed data.  For a detailed description of these plots, we refer the reader to Appendix \ref{app3}.  In the chi-square boxplots, we see the division by expected proportion in the summands of the chi-square discrepancy \eqref{chi2} reflected in the larger contribution of relative discrepancies to the reported p-values; in contrast, we see the equal-weighting of the summands of the root-mean-square distance \eqref{rms} reflected in the larger contribution of absolute discrepancies to the reported root-mean-square p-values.   In Section~\ref{analysis}, we will see that all of the classic statistics, not just the chi-square statistic, are sensitive to relative rather than absolute discrepancies.

\subsubsection{Interpretation of the results for Example 1}\label{sec4.3}

Comparing the boxplots in Figure \ref{fig:box2}, we see that both chi-square and root-mean-square tests report a significant deviation in the largest index, among others.  The largest index  corresponds to the $18$ observed counts versus $10$ expected counts of genotype $\{\textrm{A}_3,\textrm{A}_2\}$ in Example 1.  However, the p-value reported by the root-mean-square test is an order of magnitude smaller than the p-value reported by chi-square test, as this discrepancy is larger compared to expected root-mean-square fluctuations than it is compared to expected chi-square fluctuations.  In the chi-square summation, the statistical significance of this deviation (as well as the deviations in indices 6 and 7) is washed out by large expected relative deviations in the rare genotypes.

\begin{table}[ht]
\label{table1}
\caption{\baselineskip=10pt Plain and fully-conditional (FC) p-values for Pearson's statistic $X^2$, the log--likelihood-ratio statistic $G^2$, the Hellinger distance $H^2$, the negative log--likelihood statistic $L$, and the root-mean-square statistic $F$, for the observed genotypic counts in Examples 1-3  to be consistent with the Hardy-Weinberg equilibrium model \eqref{hwe1}.  With $99\%$ confidence, p-values are correct to $\pm .001$.}
\vspace{5mm}
\centering
\begin{tabular}{| l || l  l || l l || l l| }
   \hline
          &    \multicolumn{2}{|c||}{Example 1}  &  \multicolumn{2}{|c||}{Example 2}  & \multicolumn{2}{|c|}{Example 3}    \\
           \hline \
        Statistic   & Plain p-val & FC p-val & Plain p-val & FC p-val & Plain p-val & FC p-val \\ \hline

         $X^2$&  .693  & .709 &  .020  & .020  & .015 & .026   \\*[-.60em]
         $G^2$ & .600  & .630  & .013  & .013  & .181  & .276    \\*[-.60em]
          $H^2$ & .562   & .602  & .027   & .025  & .307   & .449   \\*[-.60em]
         $L$ & .648   & .714  & .016   & .018  & .155   & .207    \\*[-.60em]
         {\bf $F$} & {\bf .039} & {\bf .039} & {\bf .002  } & {\bf .002  }  & .885   & .917  \\
         \hline
         \end{tabular}
\end{table}

\subsubsection{Interpretation of the results for Example 2}

The distribution of discrepancies in Figure \ref{fig:box3} can be interpreted similarly to the boxplots from Figure \ref{fig:box2}: both the chi-square and root-mean-square tests report a {\color{BrickRed}\bf statistically} significant deviation in the 5th-largest index, corresponding to the $982$ observed counts versus $1057.6$ expected counts of genotype $\{\textrm{A}_4,\textrm{A}_1\}$ in Example 2.  However, the p-value reported by the root-mean-square test is an order of magnitude smaller than the p-value reported by chi-square test, as this discrepancy is larger compared to expected root-mean-square fluctuations than it is compared to expected chi-square fluctuations.  In the chi-square summation, the statistical significance of this deviation is washed out by large expected relative deviations in the rare genotypes.
In contrast to the $n=45$ draws from Example 1, this data set contains $n = 8297$ draws; we infer that the qualitative differences between the root-mean-square and chi-square statistic are not unique to small sample-size data.

\subsubsection{Interpretation of the results for Example 3}

Comparing the expected and observed chi-square discrepancies in Figure \ref{fig:box1}(b), we might posit that the small p-value of $.015$ that the chi-square test gives to the data in Example 3 depends strongly on the discrepancy at the 4th index on the plot, corresponding to a single draw of genotype $\{\textrm{A}_6, \textrm{A}_6\}$.  By removing this draw from the data set and re-running the chi-square goodness-of-fit test on the remaining $n=29$ draws, the chi-square statistic $X^2$ returns a p-value of $.207$, well over an order of magnitude larger  than the previous p-value, confirming that the small p-value given by the chi-square statistic for the data set in Figure 2(c) is the result of observing a single rare genotype.  The root-mean-square statistic is not as sensitive to this discrepancy.

\subsection{Power analyses}

We now compare the power and Type I error for Pearson's statistic $X^2$, the log--likelihood-ratio statistic $G^2$, the Hellinger distance $H^2$, the negative log--likelihood statistic $L$,  and the root-mean-square statistic $F$, in detecting practical deviations of genotype frequencies from those expected under HWE, namely populations with increased homozygosity (as due to inbreeding), populations with increased heterozygosity, and populations of genotypes undergoing selection \citep{chenthomson, ab97, hw08}.    The results in Table 2 support the assertion that the root-mean-square statistic and the classic statistics focus their power on complementary classes of alternatives.  In this section we will consider four parameter specifications:

\begin{enumerate}[noitemsep]
\item {\bf Alternative:} $r = 10$, $n=100$, and $\theta_1 = \theta_2 = 1/3$, and $\theta_j = 1/24$ for $3 \leq j \leq 10$
\item {\bf Alternative:} $r = 10$, $n = 100$, and $\theta_j \sim 1/j$ for $1 \leq j \leq 10$
\item  {\bf Alternative:} $r = 10$, $n = 200$,  and $\theta_1 = \theta_2 = 1/3$, and $\theta_j = 1/24$ for $3 \leq j \leq 10$
\item {\bf Alternative:}  $r = 20$, $n = 200$, and $\theta_j \sim 1/j$ for $1 \leq j \leq 20$
\end{enumerate}

\subsubsection{Deviations due to selection}

When there is selection for or against a particular allele or genotype in the population, the result is an excess or deficiency of genotypes carrying a particular allele or pair of alleles compared to what would be expected under HWE.     To account for selection, one introduces fitness parameters $w_{j,k} > 0$ into the HWE equations,
\begin{equation}
\label{select}
p_{j,k} =
\left\{ \begin{array}{cc}2 (w_{j,k}/\bar{w})\theta_j \theta_k, & 1 \leq k < j \leq r  \\
          (w_{k,k}/\bar{w}) \theta_k^2, & j = k.
  \end{array} \right.
\end{equation}
where $\bar{w}$ is a normalization constant.

We consider the scenario where the common allele $A_1$ is undergoing selection, so that genotypes carrying allele $A_1$ have higher fitness in the population:
\begin{equation}
\label{fitness}
 w_{j,k} =  \left\{ \begin{array}{ll}
1.5, &  k=1, \\
1, & \text{else.}
\end{array} \right.
\end{equation}

The power and Type I errors of the various statistical tests in detecting deviations from HWE due to selection for common alleles are listed in Table 2.   The root-mean-square statistic appears to be uniformly more powerful than the classic statistics while maintaining the correct asymptotic Type I error rate.  We will provide theoretical justification for these observations through an asymptotic analysis in Section 5.

\begin{table}[ht]
\label{table2}
\caption{\baselineskip=10pt Statistical power and Type I error of the various tests of HWE against deviations due to \underline{selection}, i.e., deviations of the form \eqref{select}  with parameters as specified in Alternatives 1-4 and fitness parameters \eqref{fitness} and deviations due to \underline{inbreeding}, i.e. deviations of the form \eqref{inbreed} with parameters as specified in Alternatives 1-4 and inbreeding parameter $f = 1/10$.  Power and Type I errors are at the $5\%$ significance level, and computed using 5000 simulations from the alternative distribution and 5000 Monte Carlo trials per each simulation.}
\vspace{5mm}

\centering
\begin{tabular}{ c || ll || ll || ll || ll}

 \multicolumn{8}{c}{Deviations due to selection for common allele}   \\
   \hline
    &  \multicolumn{2}{| c || }{Alternative 1}   &   \multicolumn{2}{|c||}{Alternative 2}  &  \multicolumn{2}{|c||}{Alternative 3}  &  \multicolumn{2}{|c}{Alternative 4}   \\
         Statistic   & Power & Type I & Power & Type I & Power & Type I & Power & Type I   \\
    \hline
         $X^2$  & .04           & {\bf .05}           &  .01               & {\bf .05}   & .04               & {\bf .05}   & $<$  .01           &{\bf  .05}   \\*[-.60em]
          $G^2$ & .07            & .06          & .02                          & .07    & .07              & .06   & .01           & .08   \\*[-.60em]
          $H^2$ & .08           & .06       &  .05                          & .07     & .08               & {\bf .05}      & .01           & .07  \\*[-.60em]
          $L$      & .03            & .04        &  .01                        & .04       & .04              & .04   & $<$ .01          & .04  \\*[-.60em]
          $F$       & ${\bf .13}$ & {\bf .05}       & ${\bf .12}$   & {\bf .05}   & ${\bf .19}$ &{\bf  .05}    & {\bf .23}    & {\bf .05}  \\
         \hline
 \multicolumn{8}{c}{Deviations due to inbreeding}   \\
   \hline
    &  \multicolumn{2}{|c|| }{Alternative 1}   &   \multicolumn{2}{|c||}{Alternative 2}  &  \multicolumn{2}{|c||}{Alternative 3}  &  \multicolumn{2}{|c}{Alternative 4}   \\
         Statistic   & Power & Type I & Power & Type I & Power & Type I & Power & Type I   \\
    \hline
         $X^2$  & .34                & {\bf .05}          &  .34             & {\bf.05}    &   .60    &.04  & .64 & .06  \\*[-.60em]
          $G^2$ & .29               & .06          & .33              & .06   &    .48    & .06 & .64  & .08  \\*[-.60em]
          $H^2$ & .18                 & .07           &  .22             & .06   &  .28   & {\bf.05}  & .42 & .07  \\*[-.60em]
          $L$      &{\bf  .39}               &{\bf .05}         & {\bf .36} & .04  &  {\bf  .63}    & .04 & {\bf.70} &  .03   \\*[-.60em]
          $F$       & .16               &{\bf  .05}       & .16                 &{\bf  .05} &  .26    & {\bf.05} & .29 & {\bf .05}  \\
         \hline
         \end{tabular}
\end{table}

\subsubsection{Deviations due to inbreeding}
We now consider genotypic distributions parameterized by an inbreeding coefficient, $f$,  which describes the extent to which members of the population with similar genetic make-up are more or less likely to mate with each other:
\begin{equation}
\label{inbreed}
p_{j,k} =
\left\{ \begin{array}{cc}2\theta_j \theta_k (1 - f), & j > k \\
          \theta_k^2 + f \theta_k (1 - \theta_k), & j = k.
  \end{array} \right. \quad \quad  1 \leq k \leq j \leq r
\end{equation}
Hardy Weinberg Equilibrium corresponds to $f = 0 $.  A negative value $f < 0$ corresponds to a deficiency of homozygotes, while a positive value of $f$ corresponds to an excess of homozygotes.
Table 2 displays the power of the various tests against alternatives of the form \eqref{inbreed} with positive inbreeding parameter.   The root-mean-square statistic appears to be less powerful than the classic statistics in detecting deviations due to inbreeding, but does appear to reach the correct asymptotic Type I error rate more quickly.

\section{An asymptotic power analysis}\label{analysis}

In this section we give theoretical justification to our assertion that the root-mean-square statistic can be more powerful than the classic statistics in detecting deviations from Hardy-Weinberg equilibrium.  To model the setting where the number of draws and number of genotypes are of the same magnitude, we consider the limit in which the number of alleles and number of draws go to infinity \emph{together}, so that the asymptotic chi-square approximation to the classic statistics is not valid in this limit. Our method is to create data sets such that the root-mean-square statistic has asymptotic power one while the chi-square statistics have asymptotic power zero.

We consider a gene having $r+1$ alleles, one common allele and $r$ rare alleles.  The \emph{Common Allele} data set we consider involves $n = 3r$ observed genotypes, distributed as indicated below.

\begin{table}[h]
\label{bigallele2}
\caption{\baselineskip=12pt Common Allele data set}
\centering
\begin{tabular}{ | l |}
   \hline
           $n=3r$ observed genotypes  \\ \hline
           $n_{1,1} = r$ of type $\{\textrm{A}_1, \textrm{A}_1\}, \quad n_{1,1}/n = 1/3$ \\
           $n_{1,k} = 2$ of type $\{ \textrm{A}_1, \textrm{A}_k\}, \quad n_{1,k}/n = 2/(3r), \hspace{1.5mm} 2 \leq k \leq r+1$ \\
           $n_{j,k} = 0$ of type $\{ \textrm{A}_j, \textrm{A}_k \}, \quad n_{j,k}/n = 0, \hspace{3mm} \quad  \quad 2 \leq j \leq k \leq r+1$  \\
           \hline
            $n_1 = 4r$ alleles of type $\textrm{A}_1, \quad n_1/(2n) = 2/3$ \\
            $n_k = 2$ \hspace{.5mm} alleles of type $\textrm{A}_k, \quad n_k/(2n) = 1/(3r), \quad 2 \leq k \leq r+1$. \\
            \hline
 \end{tabular}
\end{table}

\noindent The maximum-likelihood model counts for the Common Allele data set are
\begin{equation}
\left\{ \begin{array}{ll}
m_{1,1} = 4r/3, & \\
m_{1,k} = 4/3, & 2 \leq k \leq r+1,\\
m_{k,k} = 1/(3r), & 2 \leq k \leq r+1,  \\
m_{j,k} = 2/(3r), &  2 \leq j <  k \leq r+1, \quad j < k.
\end{array}\right.
\label{bigallele:m}
\end{equation}
To see that the Common Allele data set becomes increasingly inconsistent with the Hardy-Weinberg model as $r$ increases, observe that, under the null hypothesis, we would expect in a sample of $n=3r$ genotypes to see $r/3 = \sum_{j=2}^{r+1} \sum_{k=2}^{r+1} m_{j,k}$ genotypes containing only rare alleles. The Common Allele data set however contains \emph{no} genotypes containing only rare alleles.  In spite of this inconsistency, we will prove that the plain p-values for each of the four classic statistics $X^2, G^2$ and $H^2$, \emph{converge to 1} as $r \rightarrow \infty$, indicating zero asymptotic power.  In contrast, the p-value for the root-mean-square statistic converges to zero.

\begin{theorem}
\label{thm1}
In the limit as $r \rightarrow \infty$, the plain p-values (as computed via Algorithm \ref{alg:1} of Appendix \ref{app2} in the {\bf Supplementary Material}) given by $X^2$, the log--likelihood-ratio statistic $G^2$ and the Hellinger distance $H^2$for the Common Allele data set to be consistent with the Hardy-Weinberg equilibrium model all converge to $1$, while the plain p-value for the root-mean-square statistic converges to $0$.
\end{theorem}

The proof is given in Appendix \ref{app1}. of the {\bf Supplementary Material}.

Figure~\ref{bigallele} shows that the convergence of the classic p-values to 1, and of the root-mean-square p-value to 0, occurs very quickly.  This convergence is demonstrated for both the plain and fully conditional p-values, even though Theorem \ref{thm1} applies directly only to the plain p-values.

Finally,  the particular distribution of the draws in the Common Allele data set was somewhat arbitrary. However, a similar asymptotic analysis holds for many other data sets.  For example, we could have considered instead a data set involving two, three, or four common alleles, or one common allele and three fairly-common alleles, and so on.

\section{Concluding Remarks \label{concluding remarks}}

We have proposed the use of a simple root-mean-square statistic for testing deviations from Hardy-Weinberg equilibrium. The classic tests, tuned to detect \emph{relative} discrepancies, can be blind to overwhelmingly large discrepancies among common genotypes that are drowned out by expected finite-sample size fluctuations in rare genotypes. The root-mean-square statistic, on the other hand, is tuned to detect deviations in \emph{absolute} discrepancies, and easily detects large discrepancies in common genotypes. We demonstrated this in the analysis of three benchmark data sets of \cite{gt}, in one of which only the root-mean-square statistic gave overwhelming evidence of a departure from Hardy-Weinberg equilibrium, and another of which it gave an order of magnitude smaller p-value. We also found that the root-mean-square test can be significantly more powerful at detecting deviations from Hardy-Weinberg equilibrium arising from selection. These numerical results were complemented by the asymptotic power analysis of Section \ref{analysis}. At the very least, the root-mean-square statistic and the classic statistics focus on \emph{complementary} classes of deviations from Hardy-Weinberg equilibrium (see Figure \ref{fig:box3}), and their combined $P$-values provide a more fortified test than either $P$-value used on its own.

\baselineskip=14pt
\section*{Acknowledgment}
We thank Mark Tygert, Andrew Gelman, Abhinav Nellore, and Will Perkins for their helpful contributions.

\section*{Supplementary Material}
{\bf Supplementary Material} contains Algorithms \ref{alg:1} and \ref{alg:2} of Appendix \ref{app2} and the proof of Theorem \ref{thm1} in Appendix \ref{app1}.

\section*{Software availability}

Code for calculating plain and fully conditional p-values using the root-mean-square test statistic is available in R from http://math.utexas.edu/$\sim$rward. With appropriate citation, the code is freely available for use and can be incorporated into other programs.

\baselineskip=14pt
\bibliographystyle{biometrika}
\bibliography{HW}

\newcommand{\Appendix}{\appendix\def\thesection{Appendix~\Alph{section}}\def\thesubsection{\Alph{section}.\arabic{subsection}}}
\section*{Appendix}
\begin{figure}
\centering
{
\subfigure[]{\includegraphics[scale=.6]{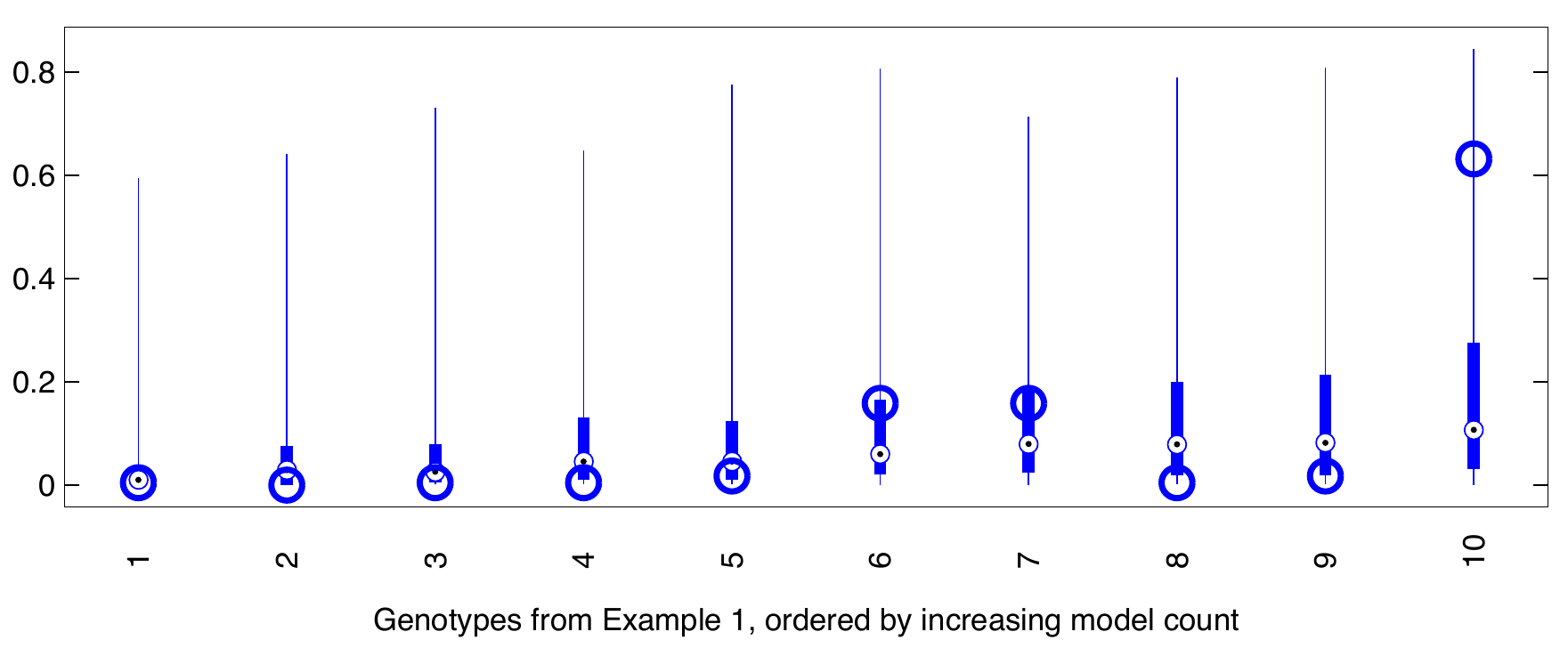}\label{fig:subfig1}} \quad \quad \quad
\subfigure[]{\includegraphics[scale=.6]{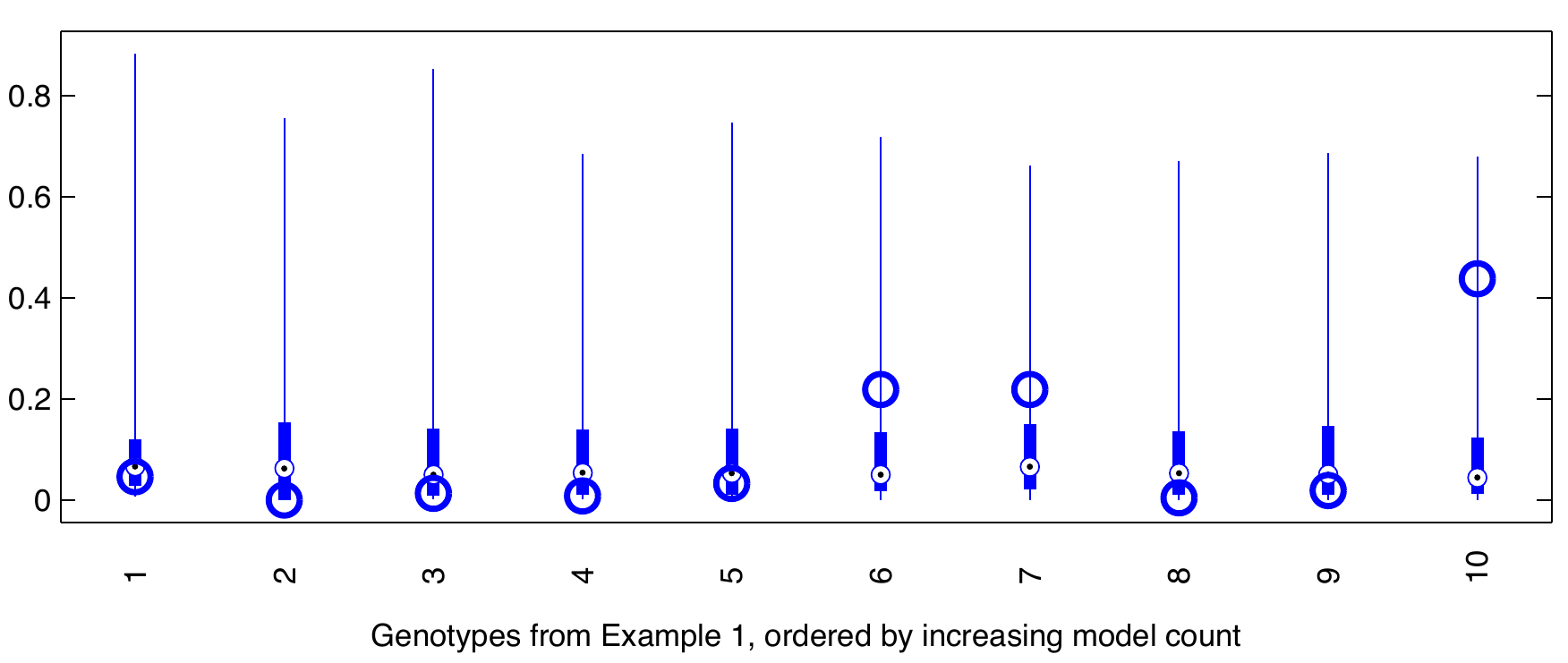}\label{fig:subfig2}}
}
\caption{\baselineskip=12pt Example 1. Expected vs. observed relative root-mean-square discrepancies (top plot) and Expected vs. observed relative $\chi^2$ discrepancies (bottom plot).}
\label{fig:box2}
\end{figure}

\begin{figure}
\centering
{
\subfigure[]{\includegraphics[scale=.8]{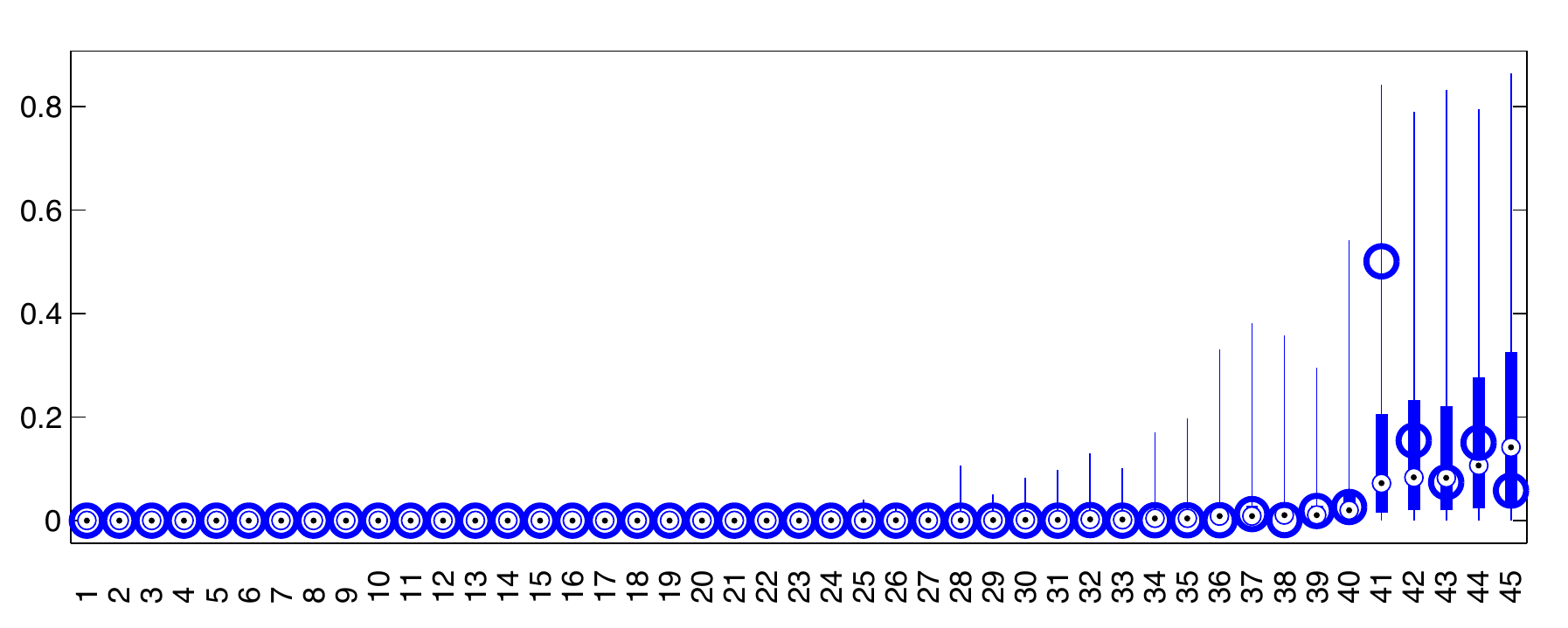}\label{fig:subfig3}} \quad \quad \quad
\subfigure[]{\includegraphics[scale=.8]{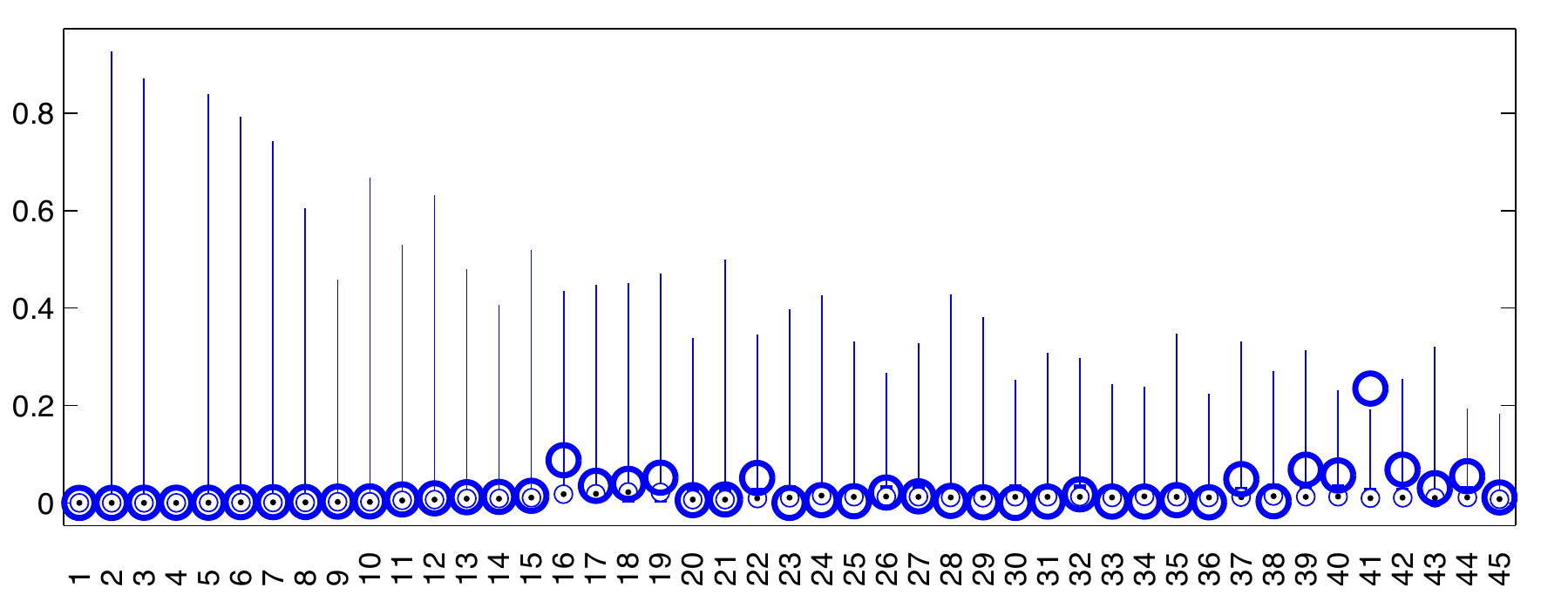}\label{fig:subfig4}}
}
\caption{\baselineskip=12pt Example 2. Expected vs. observed relative root-mean-square discrepancies (top plot) and Expected vs. observed relative $\chi^2$ discrepancies (bottom plot).}
\label{fig:box3}
\end{figure}

\begin{figure}
\centering
{
\subfigure[]{\includegraphics[scale=.8]{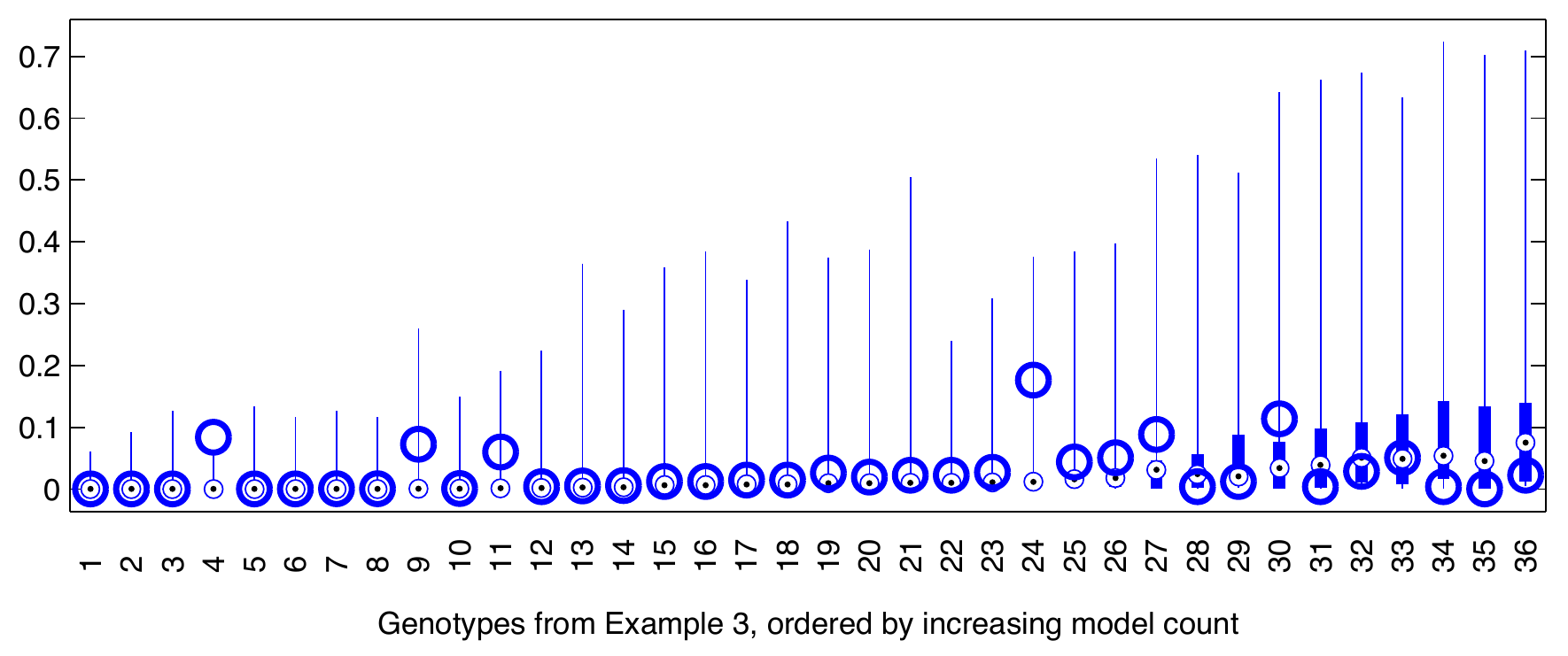}\label{fig:subfig5}} \quad \quad \quad

\subfigure[]{\includegraphics[scale=.8]{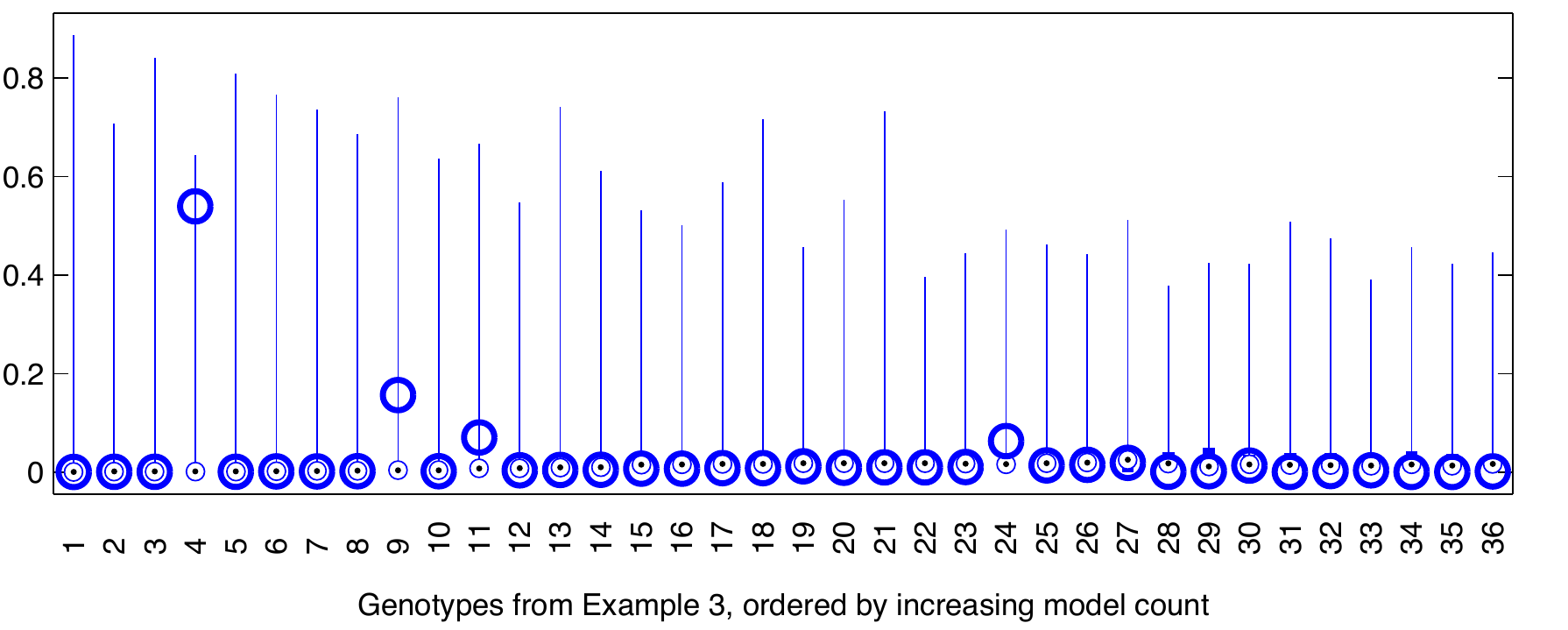}\label{fig:subfig6}}
}
\caption{\baselineskip=12pt Example 3. Expected vs. observed relative root-mean-square discrepancies (top plot) and Expected vs. observed relative $\chi^2$ discrepancies (bottom plot).}
\label{fig:box1}
\end{figure}

\begin{appendix}
\Appendix
\renewcommand{\theequation}{A.\arabic{equation}}
\renewcommand{\thesubsection}{A.\arabic{subsection}}
\setcounter{equation}{0}
\baselineskip=18pt
\subsection{Description of Figures \ref{fig:box2}, \ref{fig:box3}, and \ref{fig:box1}}\label{app3}
Consider for a sample of genotype counts the linear ordering given by the nondecreasing rearrangement of the Hardy-Weinberg equilibrium model counts: if $m_{[j]}$ denotes the $j$th smallest expected frequency among all the model genotype frequencies, $1 \leq j \leq r(r+1)/2$, then we denote the corresponding number of draws by $n_{[j]}$, and the corresponding number of observed and expected simulated draws under the (plain) HWE null hypothesis by $N_{[j]}$ and $M_{[j]}$.

The observed root-mean-square discrepancies are
\begin{equation*}
d^{rms}_j =  \big( m_{[j]} - n_{[j]} \big)^2,
\end{equation*}
while the observed chi-square discrepancies are
\begin{equation*}
d^{chi}_j =\frac{\big( m_{[j]} - n_{[j]} \big)^2}{m_{[j]}}.
\label{dchi}
\end{equation*}
The random vectors of expected root-mean-square discrepancies in $n$  i.i.d. draws from the model distribution are
\begin{equation*}
\label{Drms}
D^{rms}_j = \big( M_{[j]} - N_{[j]} \big)^2,
\end{equation*}
and
\begin{equation*}
\label{Dchi}
D^{chi}_{j} = \frac{ \big( M_{[j]} - N_{[j]} \big)^2}{M_{[j]}}.
\end{equation*}
To generate the boxplots for the relative root-mean-square discrepancies,  we simulated $K = 1000$ realizations of $n$ i.i.d. draws from the HWE model in the respective examples.   For each simulation, we computed the vector of root-mean-square discrepancies \eqref{Drms} and normalized the vector to sum to 1.  We displayed the distribution of discrepancies using a boxplot: for each term $j$, the median of the distribution $D^{(\cdot)}_j = (D^{(i)}_j)_{i=1}^{1000}$ is indicated by the bulls-eye mark $\odot$.  The rectangular box around the median extends to the $25$th and $75$th percentiles of the data, and the whiskers extending from each side of the box reach out to the $1$ and $99$th percentiles of the data. On top of the boxplot,  the observed discrepancies, $d^{rms}_j$, normalized to sum to 1, are indicated by large open circles.

The chi-square plot for each figure was created by repeating the same set-up as above using the relative chi-square discrepancies.

\end{appendix}

\clearpage\pagebreak\newpage
\pagestyle{empty}

\begin{figure}[h]
\centering
\includegraphics[scale=.8]{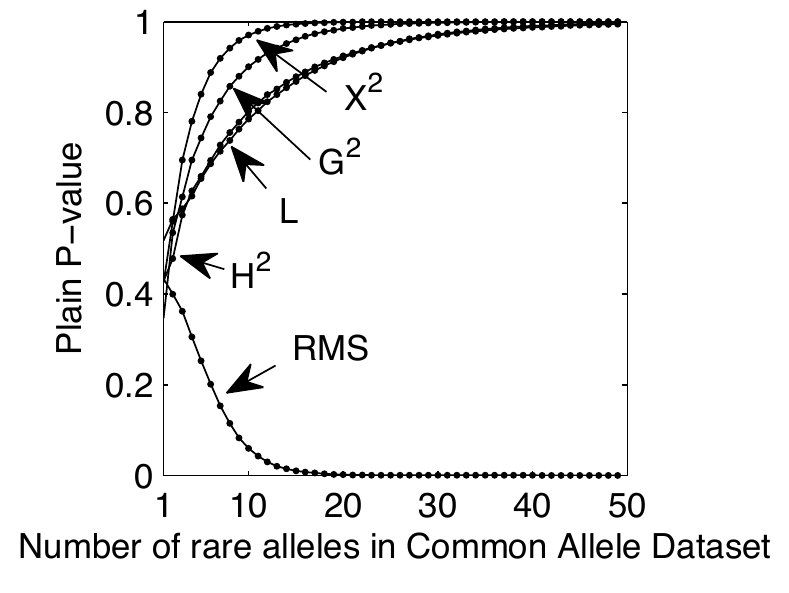}
\includegraphics[scale=.8]{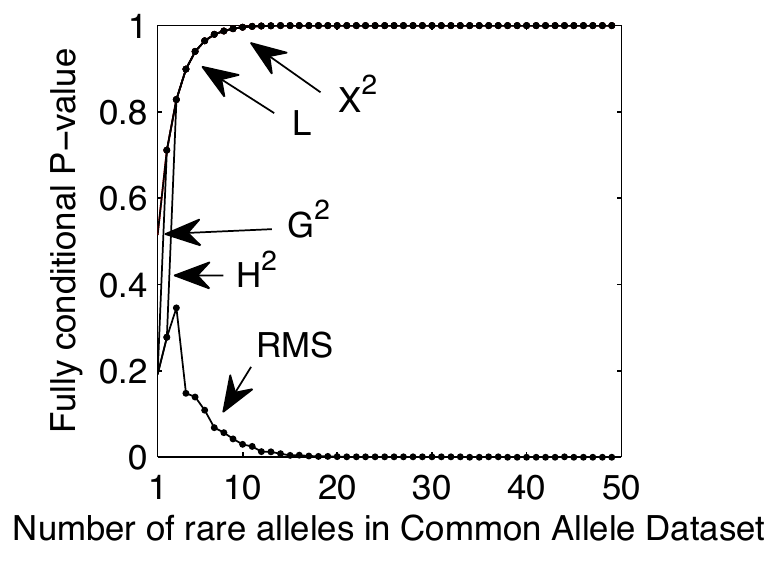}
\caption{\baselineskip=10pt  The p-values (accurate to three digits with $99\%$ confidence) for Pearson's statistic $X^2$, the log--likelihood-ratio statistic $G^2$, the Hellinger statistic $H^2$ and the root-mean-square statistic $F$ in the Common Allele data set to be consistent with the Hardy-Weinberg equilibrium model \eqref{hwe1}, as a function of the number of alleles $r$. The top plot if for the plain p-values, while the bottom plot is for the conditional p-values. }
\label{bigallele}
\end{figure}

\clearpage\pagebreak\newpage
\pagestyle{fancy}
\fancyhf{}
\rhead{\bfseries\thepage}
\lhead{\bfseries NOT FOR PUBLICATION SUPPLEMENTARY MATERIAL}
\vskip 10mm
\baselineskip=24pt

\begin{center}
{\LARGE{\bf Supplementary Material to\\ \hskip 5mm \\ {\it Testing Hardy-Weinberg equilibrium with a simple root-mean-square statistic}}}
\end{center}

\vskip 5mm
\begin{center}
Rachel Ward and
Raymond J. Carroll
\end{center}

\setcounter{equation}{0}
\setcounter{page}{1}
\setcounter{table}{1}
\setcounter{section}{0}
\renewcommand{\theequation}{S.\arabic{equation}}
\renewcommand{\thesection}{S.\arabic{section}}
\renewcommand{\thesubsection}{S.\arabic{section}.\arabic{subsection}}
\renewcommand{\thepage}{S.\arabic{page}}
\renewcommand{\thetable}{S.\arabic{table}}
\baselineskip=17pt

\pagebreak\newpage\clearpage
\section{Pseudocode for calculating exact p-values}\label{app2}

\begin{algorithm}[thb]
\caption{\baselineskip=12pt Computing the plain p-value}
	\label{alg:1}
\centering \fbox{
\begin{minipage}{.99\textwidth}
\vspace{4pt}
\alginout{Observed genotype counts $n_{j,k}$, number of Monte Carlo simulations $\ell$, and test statistic $S$ (e.g. $S = X^2, G^2, H^2, \dots )$}
{plain p-value associated to test statistic $S(n_{j,k}, m_{j,k})$}
\vspace{8pt}\hrule\vspace{8pt}

\begin{algtab*}
Compute maximum-likelihood model counts $m_{j,k} =  (2-\delta_{jk})(n_j \hspace{.5mm} n_k) / (4n)$ \\
Measure the discrepancy $s = S(n_{j,k}, m_{j,k})$. \\
	  \\		
$i \leftarrow 0$

\algrepeat
	- $i \leftarrow i + 1$
		
         - Draw $n$ genotypes $X_1^{(i)}, \dots, X_q^{(i)}, \dots, X_n^{(i)}$ i.i.d. from the multinomial model distribution \\ \quad $(m_{j,k}/n)$
 	
	- Aggregate simulated genotype counts $N^{(i)}_{j,k} =\# \big\{q : X_q^{(i)} = \{A_j, A_k \} \big\}$
	
	- Aggregate simulated allele counts $N_j^{(i)} =  \big( \sum_{k=j}^r N^{(i)}_{k,j} + \sum_{k=1}^{j} N^{(i)}_{j,k} \big)$ and proportions \\ \quad $\Theta_j^{(i)} = N_j^{(i)}/(2n)$.

        - Compute maximum-likelihood counts  $M^{(i)}_{j,k} =  (2-\delta_{jk})N^{(i)}_j N^{(i)}_k/(4n)$
 	
	- Evaluate simulated discrepancy $S_i = S(N^{(i)}_{j,k}, M^{(i)}_{j,k})$   \\
\alguntil{$i=\ell$} \\
\algreturn plain p-value, $P = \# \{i: S_{i} \geq s \} / \ell$
\end{algtab*}
\vspace{5pt}
\end{minipage}
}
\end{algorithm}

\begin{algorithm}[thb]
\caption{\baselineskip=12pt Computing the fully conditional p-value}
	\label{alg:2}
\centering \fbox{
\begin{minipage}{.99\textwidth}
\vspace{6pt}
\alginout{Observed genotype counts $n_{j,k}$ and allele counts $n_{j}$, number of Monte Carlo simulations $\ell$, and test statistic $S$ (e.g. $S = X^2, G^2, H^2, \dots )$}
{fully conditional p-value associated to test statistic $S(n_{j,k}, m_{j,k})$}
\vspace{8pt}\hrule\vspace{8pt}

\begin{algtab*}
Compute maximum-likelihood model counts $m_{j,k} = (2-\delta_{jk}) n_j n_k / (4n)$.\\
Measure the discrepancy $s = S(n_{j,k}, m_{j,k})$. \\
	  \\		
$i \leftarrow 0$

\algrepeat
	- $i \leftarrow i + 1$ \\	
	- Apply a random permutation to the sequence of alleles as in \eqref{string} to obtain $n$ simulated  genotypes $X_1^{(i)}, \dots, X_q^{(i)}, \dots, X_n^{(i)}$ with fixed allele counts $n_j$.  \\
	- Aggregate simulated genotype counts $N^{(i)}_{j,k} = \# \big\{q : X_q^{(i)} = \{A_j, A_k \} \big\}$ \\
	- Evaluate simulated discrepancy $S_i = S(N^{(i)}_{j,k}, m_{j,k})$ \\	
	\alguntil{$i=\ell$}

\algreturn	fully conditional p-value, $P = \# \{i: S_{i} \geq s \} / \ell$
\end{algtab*}
\vspace{5pt}
\end{minipage}}
\end{algorithm}

\section{Proof of Theorem 1}\label{app1}

The crux of the proof is that, as $r$ increases, relative fluctuations in the rare genotypes simulated under HWE become sufficiently large that the sum of relative discrepancies expected under the null hypothesis exceeds the sum of the observed relative discrepancies.  However, the sum of absolute fluctuations expected under the HWE model remains bounded below the sum of the observed absolute discrepancies.

In the proof of Theorem \ref{thm1}, we will use the notation $u_n \gtrsim v_n$ to indicate that there exists some absolute constant $C > 0$ such that $u_n \geq C v_n$ for all $n = \{1,2, \dots \}$.  We use the notation $u \lesssim v$ accordingly.  We will use $C>0$ to denote a positive universal constant that might be different in each occurrence.   We write $X(r) \rightarrow  y$ to mean that the distribution $X(r)$ converges to the value $y$ as $r \rightarrow \infty$.

\noindent\underline{Proof of Theorem \ref{thm1}}
Recall the relevant notation for computing plain p-values in Algorithm \ref{alg:1}, along with the Common Allele data set in Table 4 and its maximum-likelihood HWE model counts \eqref{bigallele:m}.  Here and throughout, we will refer to $A_1$ as the \emph{common} allele and to $\{A_1, A_1\}$ as the common genotype; we will refer to the remaining $r$ alleles as \emph{rare},  to genotypes of the form  $\{A_1, A_j\}, \hspace{1mm} 2 \leq j \leq r+1,$ as \emph{rare observed} genotypes, and to genotypes of the form $\{A_j, A_k\}, \hspace{1mm} 2 \leq j \leq k \leq r+1$ as \emph{unobserved} genotypes.

\begin{enumerate}
\item Because the model proportion $\theta_1 = 2/3$ remains constant as $r$ increases but the number of draws $n = 3r$ tends to infinity, the law of large numbers implies that $\Theta_1 \rightarrow \theta_1 = 2/3$. Accordingly, $M_{1,1}/n \rightarrow m_{1,1}/n = 4/9$ and $\sum_{j=2}^{r+1} \Theta_j =1 - \Theta_1 \rightarrow 1/3$.  In words, eventually $2/3$ of the simulated alleles and $4/9$ of the simulated genotypes from the model will be common.

\item Similarly,
\begin{eqnarray*}
\hbox{$\sum_{k=2}^{r+1} M_{k,1}$}/n& \rightarrow& \hbox{$\sum_{k=2}^{r+1}$} m_{1,k}/n = 4/9; \\
\hbox{$\sum_{k=2}^{r+1} \sum_{j=2}^{r+1}$} M_{k,j}/n &\rightarrow& \hbox{$\sum_{k=2}^{r+1} \sum_{j=2}^{r+1}$} m_{k,j}/n  = 1/9.
\end{eqnarray*}
In words, roughly $4/9$ of the draws simulated from the model will be \emph{rare observed} genotypes, while $1/9$ of the simulated draws will \emph{unobserved} genotypes.

\item With probability approaching 1 as $r \rightarrow \infty$, each of the roughly $n/9 = r/3$ simulated draws from the pool of $(r^2 - r)/2$ unobserved genotypes will have a different genotype from the others.  At this point, roughly $r/3$ of the unobserved simulated proportions $N_{j,k}/n$, $2 \leq k \leq j \leq r+1$, will equal $1/(3r)$, while the others will equal 0.

\item The coupon collector's problem (see, for example, \cite{mr95}) implies that with probability approaching $1$ as $r \rightarrow \infty$, among the roughly $2r$ simulated draws from the pool of $r$ rare alleles, no rare allele will be drawn more than $\log(r)$ times (fixing the base of the logarithm at any real number greater than 1 that does not depend on $r$), and at least $3r/4$ among the $r$ rare alleles will be drawn at least twice.
\end{enumerate}
In particular, the last point above implies that, with probability approaching 1 as $r \rightarrow \infty$, all of the simulated rare proportions $\Theta_j = \Theta_j(r), \hspace{1mm} 2 \leq j \leq r+1,$ will satisfy

\begin{equation}
\label{theta1}
\Theta_j(r) \leq \log(r)/r
\end{equation}
and, for at least $3r/4$ among the $r$ simulated rare proportions,
\begin{equation}
\label{theta2}
1/(3r) \leq \Theta_j(r) \leq \log(r)/r.
\end{equation}

\begin{enumerate}
\item {\bf The p-value for the root-mean-square goes to 0 when $r \rightarrow \infty$.} The measured sum-square discrepancy $\widetilde{f}^2 = r(r+1) f^2/2$ between the observed proportions $n_{j,k}/n$ and the model proportions $m_{j,k}/n$ is
\begin{eqnarray*}
\widetilde{f}^2 &=&   \Big( \frac{n_{1,1}}{n} - \frac{m_{1,1}}{n} \Big)^2 + \sum_{k=2}^{r+1} \Big( \frac{n_{k,1}}{n} - \frac{m_{k,1}}{n} \Big)^2 + \sum_{2 \leq k \leq j \leq r+1} \Big( \frac{m_{j,k}}{n} \Big)^2  \nonumber \\
&=&  \Big( \frac{1}{9} \Big)^2 + \frac{4}{81r}+\frac{1}{81r^3} +\frac{2(r-1)}{81r^3}.
\end{eqnarray*}
As $r \rightarrow \infty$,
\begin{equation}
\label{rmstoinf}
\widetilde{f} \rightarrow 1/9.
\end{equation}
If we instead consider the sum-square statistic $\widetilde{F}^2 = \frac{(r+1)(r+2)}{2} F^2$ resulting from drawing $n=3r$ genotypes i.i.d. from the model distribution \eqref{bigallele:m}, points 1, 3, and 4 above give
\begin{eqnarray}
\label{rmstoinf2}
\widetilde{F}^2 &\lesssim& \frac{(N_{1,1} - 4r/3)^2}{9r^2}+ \sum_{k=2}^{r+1} \Big(\frac{(\log{r})^2}{r} \Big)^2 \nonumber \\
&+& \sum_{2 \leq k \leq j \leq r+1: N_{j,k}=1} \Big(\frac{1}{3r} \Big)^2 +  \sum_{2 \leq k \leq j \leq r+1: N_{j,k}=0} \Big(\frac{\log{r}}{r} \Big)^4 \nonumber \\
&\sim& \frac{Z^2}{27r/4} + \frac{(\log{r})^4}{r} + \frac{r}{3} \frac{1}{9r^2} + \Big( \frac{r(r+1)}{2} - \frac{r}{3} \Big) \Big(\frac{\log{r}}{r} \Big)^4,
\end{eqnarray}
where $Z = (N_{1,1} - 4r/3)/\sqrt{4r/3}$ converges in distribution to a standard normal distribution as $r \rightarrow \infty$.  Therefore, as $r \rightarrow \infty$,
$\widetilde{F} \rightarrow 0.$ Combining \eqref{rmstoinf} and \eqref{rmstoinf2} shows that the p-value for the root-mean-square statistic, $P = \pr\{F \geq f \} = \pr \{ \widetilde{F} \geq \widetilde{f} \}$, goes to $0$ as $r \rightarrow \infty$.

\item {\bf The p-value for $X^2$ goes to 1 as $r \rightarrow \infty$.} Similar to the measured sum-square discrepancy $\widetilde{f}$, the measured $\chi^2$ discrepancy $\tilde{\chi}^2  = \chi^2/n$ converges to some finite positive real number as $r \rightarrow \infty$.  Alternatively, if we simulate $n=3r$ genotypes from the model distribution and (following point 3 above) consider only those roughly $r/3$ summands in the normalized $\chi^2$ statistic $\tilde{X}^2 =  X^2/n$ corresponding to the unobserved genotypes with one simulated draw,
\begin{eqnarray}
\tilde{X}^2 &\gtrsim&  \frac{r}{3} \min_{2 \leq k \leq j \leq r + 1: N_{j,k} = 1}\Big(\frac{N_{j,k}}{n} - \frac{M_{j,k}}{n} \Big)^2 / \Big(\frac{M_{j,k}}{n} \Big)\nonumber \\
& \gtrsim& \frac{r}{3} \Big(\frac{1}{3r} - \Big( \frac{\log{r}}{r} \Big)^2 \Big)^2 / \Big( \frac{\log{r}}{r} \Big)^2. \label{eqa6}
\end{eqnarray}
It follows from (\ref{eqa6}) that  $\tilde{X}^2 \gtrsim r / \{\log(r)\}^2 \rightarrow \infty,$ and so the p-value for the $\chi^2$ statistic, $P = \pr (X^2 \geq \chi^2) = \pr (\tilde{X}^2 \geq \tilde{\chi^2}^2)$, goes to $1$ as $r \rightarrow \infty$.

\item {\bf  The p-value for the Hellinger statistic $H^2$ goes to 1 when $r \rightarrow \infty$.}
We have to be a bit more careful with the analysis of the Hellinger discrepancy $\tilde{h}^2 = h^2/(4n)$.
The observed discrepancy is
\begin{eqnarray}
\label{littleh}
\tilde{h}^2 &=& \frac{(\sqrt{3}-2)^2}{9} + \sum_{j=2}^{r+1} \Big(\sqrt{\frac{2}{3r}} -  \sqrt{\frac{4}{9r}}  \Big)^2 + \sum_{2 \leq k < j \leq r+1}  \frac{2}{9r^2} + \sum_{j=2}^{r+1} \frac{1}{9r^2} \nonumber \\
&=& \frac{(\sqrt{3}-2)^2}{9} + \frac{10 - 4\sqrt{6}}{9} + \frac{1}{9} \nonumber \\
&=& .14....
\end{eqnarray}

Alternatively, suppose we simulate $n=3r$ genotypes from the model distribution and consider $r$ sufficiently large.   Each estimated rare allele proportion will be bounded: $\Theta_j \leq \log(r)/r$, as stated in \eqref{theta1}
.  Furthermore, by \eqref{theta2}, at least $3/4$ of these proportions will satisfy $\Theta_j \geq 1/(3r)$, ensuring that at least $\frac{(3/4)^2 r^2}{2} - r$ among the $r(r+1)/2$ simulated proportions for the unobserved genotypes satisfy $M_{j,k}/n \geq 2/(9r^2)$.  Then, for sufficiently large $r$,
 \begin{eqnarray}
 \label{bigh}
\tilde{H}^2 &\geq& \sum_{2 \leq j \leq k \leq r+1} \Big( \sqrt{N_{j,k}/n} - \sqrt{M_{j,k}/n} \Big)^2\nonumber \\
&\geq& \# \{ j,k : N_{j,k} = 1 \} \Big(  \frac{1}{\sqrt{3r}} - \frac{\log{(r)}}{r} \Big)^2 \nonumber \\
&+& \Big(  \Big(\frac{3}{4} \Big)^2\frac{r^2}{2} - r - \# \{ j,k: N_{j,k} = 1 \} \Big) \Big( \frac{2}{9r^2} \Big) \nonumber \\
&\sim& \frac{r}{3} \Big( \frac{1}{\sqrt{3r}} - \frac{\log{r}}{r} \Big)^2 + \Big( \Big(\frac{3}{4} \Big)^2 \frac{r^2}{2} - r - \frac{r}{3} \Big) \Big(\frac{2}{9r^2} \Big) \nonumber \\
&\rightarrow & .17 ....
\end{eqnarray}
Combining \eqref{littleh} and \eqref{bigh}, we conclude that the p-value for the Hellinger distance, \\
$P = \pr( H^2 \geq h^2 )=\pr( \tilde{H}^2 \geq \tilde{h}^2 )$, goes to $1$ as $r \rightarrow \infty$.
\end{enumerate}

\end{document}